# MEMS-Reconfigurable Metamaterials and Antenna Applications

Tomislav Debogovic and Julien Perruisseau-Carrier

*Abstract*—This paper reviews some of our contributions to reconfigurable metamaterials, where dynamic control is enabled by micro-electro-mechanical systems (MEMS) technology. First, we show reconfigurable composite right/left handed transmission lines (CRLH-TLs) having state of the art phase velocity variation and loss, thereby enabling efficient reconfigurable phase shifters and leaky-wave antennas (LWA). Second, we present very low loss metasurface designs with reconfigurable reflection properties, applicable in reflectarrays and partially reflective surface (PRS) antennas. All the presented devices have been fabricated and experimentally validated. They operate in X- and Ku-bands.

## I. INTRODUCTION

Metamaterials [1]-[5] allow efficient manipulation of guided and free-space electromagnetic waves, thereby potentially improving existing and enabling novel microwave component and antenna designs [6]-[11]. On the other hand, dynamic reconfiguration [12]-[17] has become a prime need in modern communication and sensing systems, for instance to scan space and polarization, to dynamically compensate for varying system conditions thereby guaranteeing optimal performance in real time, or to support a higher number of functionalities through a single and compact device. In this context the implementation of reconfigurable metamaterials [18], [19] for antenna applications has become a topic of intense practical relevance. For example, reconfigurable metamaterials can enable operating frequency reconfiguration [13], [20], beam steering without the need for a beam-forming network [16], [21], and dynamic beamwidth control [22], [23].

The technology used to enable reconfiguration has a profound impact on the performance of reconfigurable metamaterials and consequently on the quality of microwave components and antennas utilizing them. Micro-electro-mechanical systems (MEMS) technology [24]-[30] can provide excellent properties thanks to very low loss, virtually zero power consumption (electrostatic control), high linearity, and possibility of monolithic integration.

In this paper we show reconfigurable metamaterial devices with direct applications in antenna systems. In Section II we present a Ku-band MEMS-reconfigurable composite right/left handed transmission lines (CRLH-TLs) with applications in leaky-wave antennas (LWA) and feed networks for antenna arrays. In Section III we present X-band MEMS-reconfigurable metasurfaces whose reflection properties can be reconfigured. These devices can be applied for dynamic beam-scanning/forming in reflectarrays and dynamic beamwidth control in partially reflective surface (PRS) antennas. The underlying reconfiguration technology is MEMS due to the need for low loss, convenient control, and ability to monolithically integrate a large number of controlling elements into an electrically large structure. Finally, conclusions are drawn in Section IV.

## II. MEMS-RECONFIGURABLE CRLH-TLS

A composite CRLH-TL structure [4], [5], [31]-[33] is a class of 1-D metamaterial, which can be implemented by lumped 'dual' elements (series capacitors $C_s$ and shunt inductors $L_p$) loading a usual TL, as shown in Fig. 1a. It is interesting because it can provide both positive and negative phase shift, corresponding to the left-handed and right-handed region of the Bloch-Floquet propagation constant, respectively, with seamless transition between them. In addition, it is inherently wideband structure, thereby being well suited for low loss phase shifting and antenna applications. MEMS technology enables relatively straightforward phase shift reconfiguration by dynamically controlling the series capacitance $C_s$ [34], [35] or both series capacitance $C_s$ and shunt inductance $L_p$ [36].

One possible unit cell design, implementing the circuit model of Fig. 1a, is shown in Fig. 1b [34]. A similarly operating design can be found in [35] as well. Series MEMS capacitors and folded shunt stub inductors load a coplanar waveguide (CPW) line. The profile of the MEMS area is shown in the inset of Fig. 1b. The bottom MEMS electrode is a part of the CPW signal line in the middle of the unit cell, and it is dc-grounded through inductive stubs. The movable MEMS

This work was supported by the Swiss National Science Foundation (SNSF) under grant n°133583.

T. Debogovic is currently with the Laboratory of Electromagnetics and Acoustics, Ecole Polytechnique Fédérale de Lausanne (EPFL). Mailing address: EPFL-STI-IEL-LEMA, ELB 030 (Bât. ELB), Station 11, CH-1015 Lausanne, Switzerland. E.mail: tomislav.debogovic@epfl.ch.

J. Perruisseau-Carrier is with the group for Adaptive MicroNanoWave Systems, Ecole Polytechnique Fédérale de Lausanne (EPFL). Mailing address: EPFL-STI-IEL-GR-JPC, ELB 030 (Bât. ELB), Station 11, CH-1015 Lausanne, Switzerland. E-mail: julien.perruisseau-carrier@epfl.ch.

membrane, essentially operating as the series capacitor $C_s$, is connected to the CPW signal line at the unit cell input/output by the main anchor. Electrostatic MEMS actuation, enabling analog capacitance control, is achieved by applying a voltage between CPW signal line and grounds at the cell input/output. Several unit cells can be cascaded by using dc-block capacitors and high resistivity bias lines. The unit cell was fabricated using MEMS process developed at Middle East Technical University on 500 µm-thick glass wafers, detailed in [27].

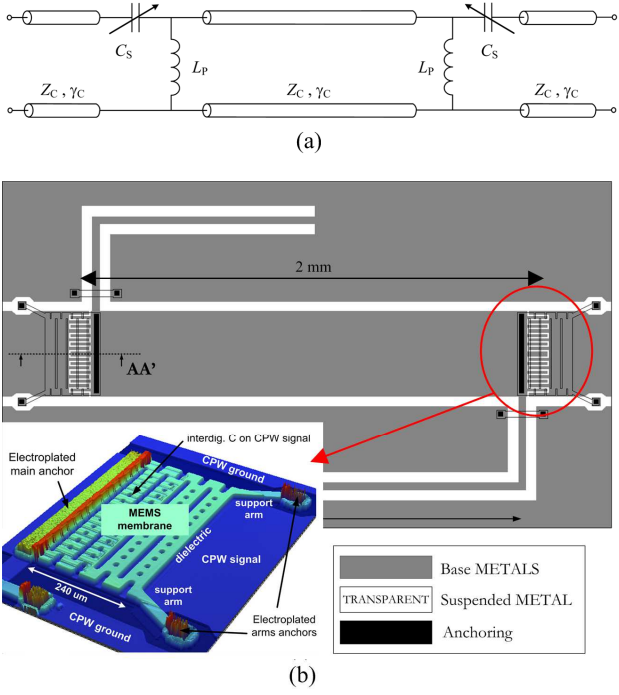

Fig. 1 MEMS-based CRLH-TL unit cell from [34]: (a) Circuit model; (b) Layout.

Simulated propagation constant $\gamma_B$ (without MEMS actuation) is shown in Fig. 2a. A typical CRLH-TL dispersion is observed, with the frequency of the 0° phase shift (corresponding to the transition between the left- and right-handed bands) at $f_0 = 14$ GHz. Simulated and measured results of the transmission phase upon reconfiguration are shown in Fig. 2b. Reconfigurable CRLH-TLs have at least two applications in antennas, both based on the transmission phase manipulation shown in Fig. 2b. The first one concerns leaky wave antennas scanning around broadside [4], [5], [7] and it relies on the ability to generate a reconfigurable negative/zero/positive phase shift at a given operating frequency, as symbolized by the 'A' arrow in the figure. The second main application concerns series feed networks or dividers [4], [5]. These devices operate at the frequency of 0° phase shift. Thus, their frequency of 0° phase shift $f_0$, in fact their operating frequency, could be reconfigured as shown by the 'B' arrow in Fig. 2b. Owing to the MEMS technology, very low insertion loss is achieved, being less than 0.7 dB when the transmission phase is reconfigured, and less than 0.8 dB when the frequency of 0° phase shift is reconfigured. A differential phase shift over losses is 38°/dB at 14 GHz.

Apart from the analog MEMS control presented above, digital MEMS control (where MEMS elements take two discrete states, namely 'up' and 'down') can also be implemented. In fact, this type of control is less sensitive to MEMS fabrication tolerances and vibrations, and nowadays it is widely accepted. A CRLH-TL unit cell with digital MEMS control [36] is shown in Fig. 3. A digital MEMS capacitor located in the horizontal CPW line acts as the series capacitor, while the remaining two (identical) MEMS elements, located in the vertical shorted CPW stubs serve to effectively change their length and hence the shunt inductance value. As a result, a CRLH-TL line with the phase shift of −50°/+50° is obtained.

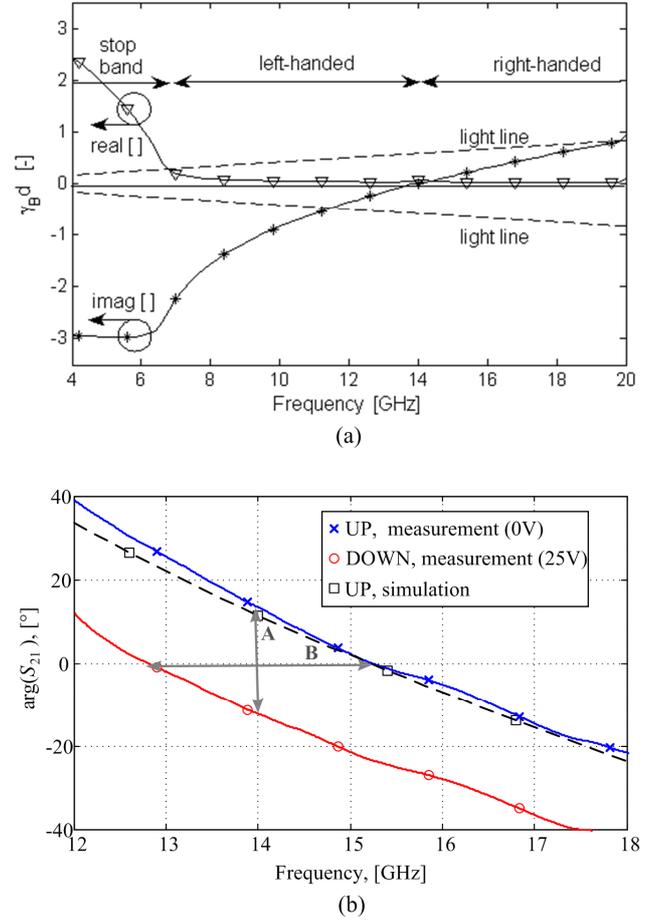

Fig. 2 MEMS CRLH-TL unit cell results from [34]: (a) Simulated Bloch-Floquet propagation constant without MEMS actuation; (b) Simulated and measured transmission phase upon reconfiguration. Arrows 'A' and 'B' illustrate possible modes of utilization.

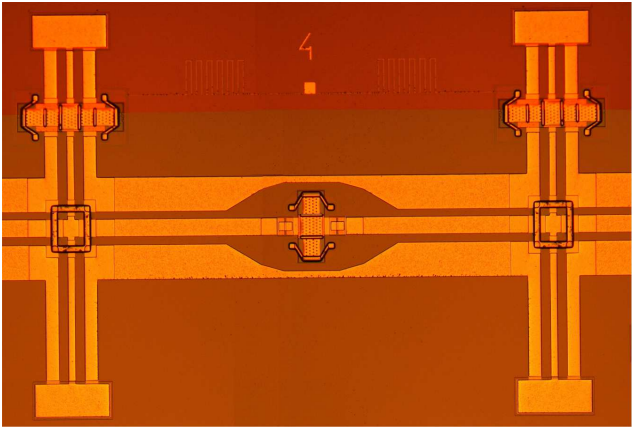

Fig. 3 Digitally controlled MEMS-based CRLH-TL unit cell from [36]. Series capacitance and shunt inductances are controlled in two discrete states.

## III. MEMS-Reconfigurable Metasurfaces

This section presents unit cells of two MEMS-reconfigurable metasurface types. The first type is designed to provide *reflection phase* reconfiguration and it has direct application in reconfigurable reflectarray antennas [17], where shaping of reflection phase profile of a reflector allows dynamic beam-scanning. Such functionality and essentially flat reflectarray antenna geometry are very useful in space communications and radars.

The second metasurface type presented is a partially reflective (or a partially transparent) design whose *reflection magnitude* can be reconfigured owing to the embedded MEMS elements. Placing such a metasurface approximately half a wavelength above a source antenna backed by a ground plane results in a reconfigurable partially reflective surface (PRS) antenna [37], whose directivity (beamwidth) depends on the metasurface reflectivity. Therefore, directivity and beamwidth of such an antenna can be dynamically controlled by the MEMS elements. This functionality is useful for instance in satellite communications from elliptical orbits, where the antenna coverage should remain constant despite variable platform altitude.

### A. Metasurface unit cell with reflection phase control

Fig. 4 shows the reconfigurable reflection phase metasurface unit cell [38]. In essence, it is a tunable resonator that consists of two pseudo-rings backed by a ground plane, and loaded by digital series MEMS capacitors. As the structure is backed by the ground plane, its reflection loss is only a subject to the dissipation in the materials and MEMS elements, being generally very low. The MEMS elements are organized in five pairs in order to retain symmetry and thereby lower the cross-polarization level. The resonance is tuned by altering the capacitance values of the MEMS pairs. This implies that the metasurface reflection phase can be reconfigured in $2^5 = 32$ discrete states at a given operating frequency.

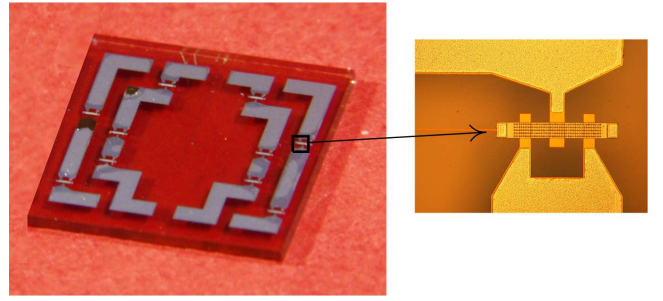

Fig. 4 MEMS-reconfigurable metasurface unit cell with reflection phase control [38]. Ten digitally controlled MEMS elements are organized in five pairs.

The metasurface reflection phase values for all 32 combinations of the MEMS states is shown in Fig. 5. It can be noticed that the reflection phase can be reconfigured within full 360° range. In addition, the phase curves at the limits of 5% bandwidth are quasi-parallel to the one observed at the operating frequency, as required to avoid beam squint in a reconfigurable reflectarray [39]. The observed reflection loss is very low, mainly below 0.3 dB.

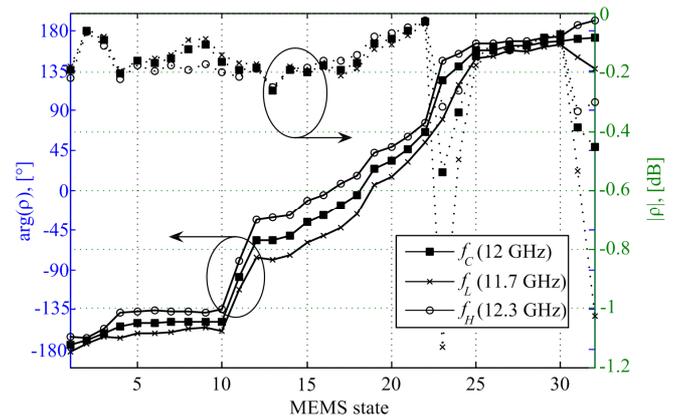

Fig. 5 Simulation results from [38]: metasurface reflection phase and loss as a function of the MEMS state, at the limit of a 5% bandwidth at 12 GHz.

The main design effort in such metasurfaces is to obtain the uniform phase step between MEMS states, and to retain it in a given bandwidth, while also keeping the phase range close to 360° and loss as small as possible. This is a matter of a compromise, but it can be nearly achieved by employing an optimization routine developed in [40], where full-wave simulation results are combined with post-processing data [41] in order to fulfil the desired goal.

The fabrication process of this unit cell is based on the deposition of a 500-nm-thick gold layer and a 1500-nm-thick aluminum one on a quartz substrate [42]. It includes the possibility to pattern high-resistivity polysilicon bias lines which do not impact on the microwave performance, which is a very important feature when designing MEMS-based metamaterials.

### B. Metasurface unit cell with reflection magnitude control

Fig. 6 shows the layout and photography of the reconfigurable reflection magnitude metasurface unit cell [43].

The unit cell topology is based on a ring-like shape, loaded by digital series MEMS capacitors $S_x$ and $S_y$. The MEMS elements come by pairs to preserve symmetry. Owing to this property, the MEMS pairs $S_x$ and $S_y$ affect reflection properties independently in $x$- and $y$-polarizations, respectively, thereby enabling a 1-bit dual-polarization operation. This unit cell is fabricated by RF Microtech using the process described in [44]. It also includes the possibility to pattern high-resistivity polysilicon bias lines which do not impact on the microwave performance (visible in Fig. 6a as narrow lines).

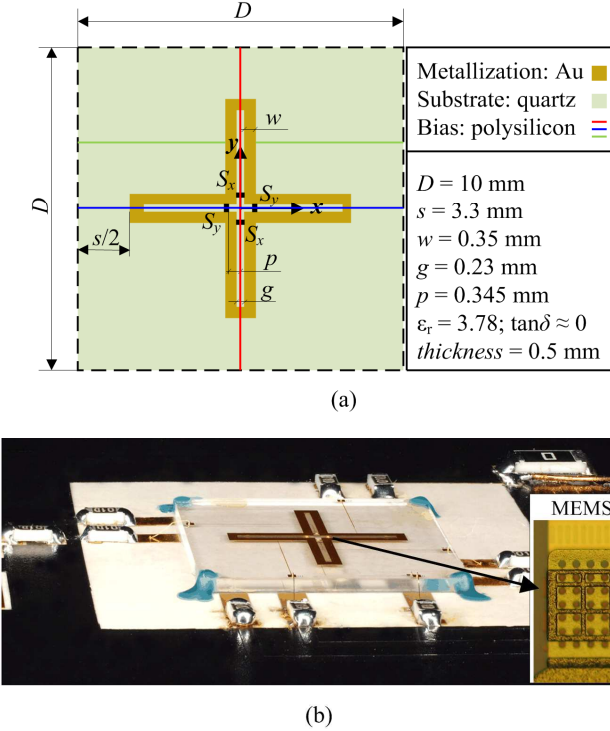

Fig. 6 MEMS-reconfigurable metasurface unit cell with reflection magnitude control from [43]: (a) Layout; (b) Photography of the fabricated device.

The unit cell reflection coefficient is shown in Fig. 7a, for incident $x$-polarization at the design frequency $f_0 = 11.2$ GHz. The reflection magnitude in this polarization is controlled (in two discrete states) by the MEMS pair $S_x$ and independent of the states of the MEMS pair $S_y$. As the metasurface is practically lossless and single-layered, the reflection magnitude reconfiguration implies a reflection phase variation as well. However, in a properly assembled reconfigurable PRS antenna, the amplitude reconfiguration has a stronger impact [23].

Predicted radiation patterns of the assembled PRS antenna, formed by placing such a metasurface above a source antenna (and thereby creating a Fabry-Pérot resonator) is shown in Fig. 7b. It can be noticed that the antenna beamwidth is reconfigured in two discrete states by MEMS elements. Since the result is shown for the $x$-polarization, only the MEMS elements $S_x$ affect the beamwidth. Thanks to the MEMS-technology and the unit cell design, the predicted radiation efficiency is above 75%, which is considered to be very high for a reconfigurable antenna operating in X-band. Results for the $y$-polarization are not shown here for space consideration, but they are essentially the same owing to the unit cell symmetry.

Reflection measurement results [43] are shown in Fig. 8. An orthomode transducer (OMT) was employed for this measurement. Being a square waveguide with couplers designed to separate its orthogonally polarized modes, the OMT allows validation of the unit cell reflectivity in two orthogonal linear polarizations. A very good agreement between measurements and simulations is observed, both in magnitude and phase, thereby validating the unit cell design and confirming the predicted reconfigurable PRS antenna performance.

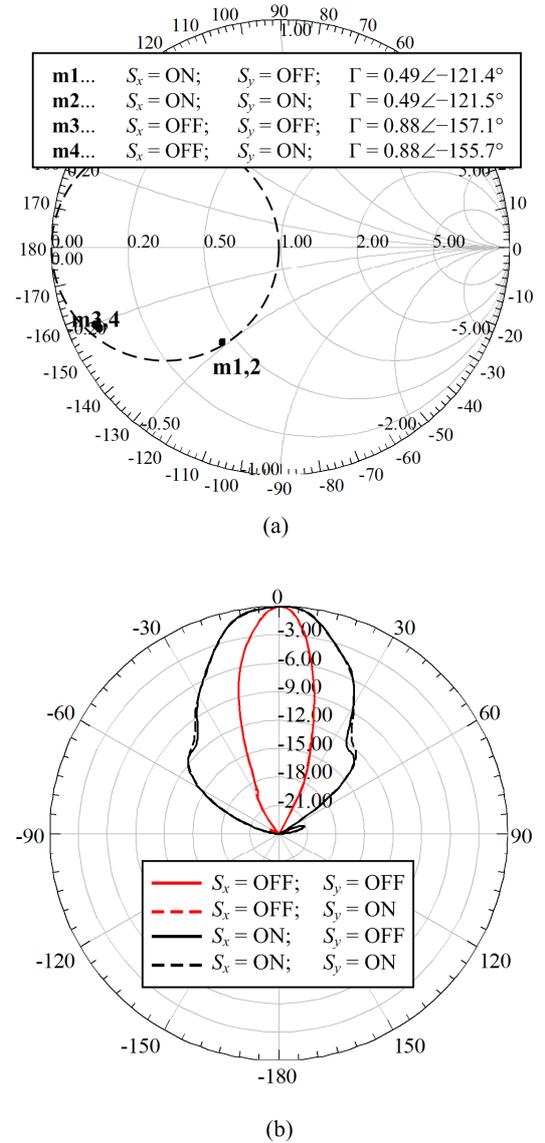

Fig. 7 Simulation results from [43]: (a) Metasurface reflection coefficient as a function of the MEMS state at the operating frequency $f_0 = 11.2$ GHz; (b) PRS antenna beamwidth as a function of the MEMS state.

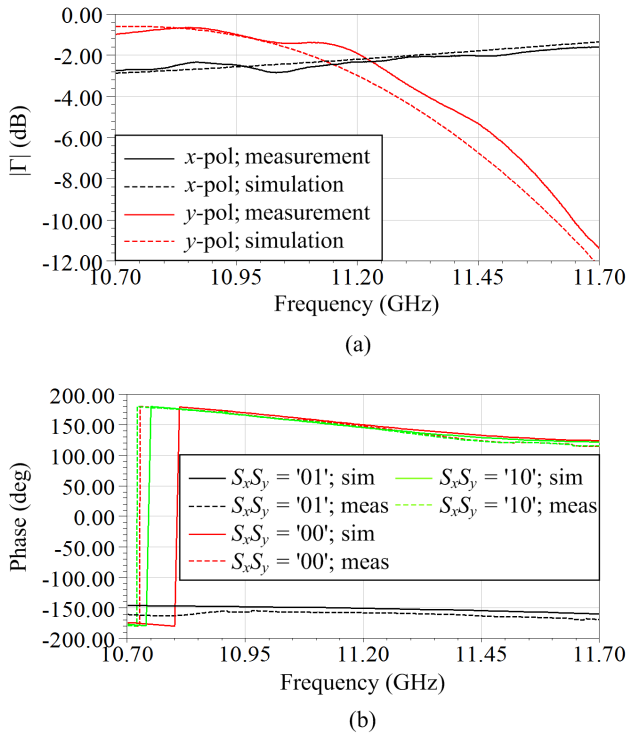

Fig. 8 Simulation and measurement results in the OMT environment, from [43]: (a) Metasurface reflection magnitude (b) Metasurface reflection phase.

The main design effort concerning such metasurfaces is to obtain a considerable reflection magnitude range while keeping the loss as low as possible. This can be achieved at the unit cell level. Then, owing to the modelling procedure developed by authors [45], the whole PRS antenna simulation can be performed accurately and rapidly despite to potentially large electrical size of the metasurface.

## IV. CONCLUSION

Monolithic MEMS integration enables structures having a potentially large number of embedded MEMS elements, at a price that does not depend directly on their number. This property, combined with other advantages of MEMS technology, such as extremely low power consumption, linearity, and availability of highly resistive bias lines, makes large reconfigurable 1-D and 2-D metamaterial designs with hundreds or even thousands of MEMS elements feasible. The main limitation of this technology lies in the actuation speed which is typically in the range of microseconds.

MEMS-reconfigurable metamaterials, if properly designed, can bring valuable improvements to antenna systems, which was demonstrated in this paper. In Section II, efficient MEMS-based CRLH-TLs were presented. It was shown that they can enable leaky-wave antennas with dynamic beam scanning and improved array feed networks In Section III, MEMS-based metasurface designs with reconfigurable reflection properties were presented. Their utilization in reflectarrays and PRS antennas, resulting in dynamic radiation pattern control (beam scanning and beamwidth reconfiguration), was discussed and demonstrated. These functionalities are welcome in radar systems and space communications.

Current trends within the area of reconfigurable metamaterials for antenna applications are mainly oriented toward higher operating frequencies. MEMS technology can already be considered as a mature technology, with prominent properties up to mm-wave frequencies. Emerging new technologies such as electrically actuated elastomers [46], [47] liquid crystals [48], [49] and graphene [50], [51], promise good reconfigurable antenna solutions at higher frequencies, up to terahertz bands.